\documentclass[twocolumn,preprintnumbers,amsmath,aps]{revtex4}
\usepackage{graphicx}
\usepackage{dcolumn}
\usepackage{bm}
\usepackage{color}
\usepackage{multirow}
\usepackage{epstopdf}
\usepackage[T1]{fontenc}
\begin{document}
\newcommand{\kvec}{\mbox{{\scriptsize {\bf k}}}}
\newcommand{\qvec}{\mbox{{\scriptsize {\bf q}}}}
\def\eq#1{Eq.\hspace{1mm}(\ref{#1})}
\def\fig#1{Fig.\hspace{1mm}\ref{#1}}
\def\tab#1{Tab.\hspace{1mm}\ref{#1}}
\title{From $\rm LaH_{10}$ to room--temperature superconductors}
\author{M. Kostrzewa $^{\left(1\right)}$}
\author{K. M. Szcz{\c{e}}{\'s}niak$^{\left(2\right)}$}
\author{A. P. Durajski $^{\left(3\right)}$}
\author{R. Szcz{\c{e}}{\'s}niak $^{\left(1,3\right)}$}
\affiliation{$^{\left(1\right)}$ Institute of Physics, Jan D{\l}ugosz University in Cz{\c{e}}stochowa, 
             Ave. Armii Krajowej 13/15, 42-200 Cz{\c{e}}stochowa, Poland}
\affiliation{$^{\left(2\right)}$ Faculty of Chemistry, University of Warsaw, Pasteura 1, 02-093 Warsaw, Poland}    
\affiliation{$^{\left(3\right)}$ Institute of Physics, Cz{\c{e}}stochowa University of Technology, 
             Ave. Armii Krajowej 19, 42-200 Cz{\c{e}}stochowa, Poland}
\date{\today} 
\begin{abstract}
Thermodynamic parameters of the $\rm LaH_{10}$ superconductor were an object of our interest. $\rm LaH_{10}$ is characterised by the highest experimentally observed value of the critical temperature: $T^{a}_{C}=215$~K ($p_{a}=150$~GPa) and $T^{b}_{C}=260$~K ($p_{b}=190$~GPa). It belongs to the group of superconductors with a~strong electron--phonon coupling ($\lambda_{a}\sim 2.2$ and $\lambda_{b}\sim 2.8$). We calculated thermodynamical parameters of this superconductor and found that the values of the order parameter, the thermodynamic critical field, and the specific heat differ significantly from the values predicted by the conventional BCS theory.

Due to the specific structure of the Eliashberg function for the hydrogenated compounds, the qualitative analysis suggests that the superconductors of the ${\rm La_{\delta}X_{1-\delta}H_{10}}$--type (LaXH--type) structure, where ${\rm X}\in\{{\rm Sc},{\rm Y}\}$, would exhibit significantly higher critical temperature than $T_{C}$ obtained for $\rm LaH_{10}$.
In the case of LaScH we came to the following assessments: $T^{a}_{C}\in\left<220,267\right>$~K and $T^{b}_{C}\in\left<263,294\right>$~K, while the results for LaYH were: $T^{a}_{C}\in\left<218,247\right>$~K and $T^{b}_{C}\in\left<261,274\right>$~K.    

\vspace*{0.25cm}
\end{abstract}
\maketitle

 The experimental discovery of the high--temperature superconducting state in the compressed hydrogen and sulfur systems 
${\rm H_{2}S}$ ($T_{C}=150$~K for $p=150$~GPa) and ${\rm H_{3}S}$ ($T_{C}=203$~K for $p=150$~GPa) \cite{Drozdov2014A, Drozdov2015A} accounts for carrying out investigations, which can potentially lead to the discovery of a material showing the superconducting properties at room temperature.

For the first time, the possibility of existence of the superconducting state in hydrogenated compounds was pointed out by Ashcroft in 2004 \cite{Ashcroft2004A}. It was stated in his second fundamental work concerning the high--temperature superconductivity, following his first work written in 1968, in which he propounded the existence of the high--temperature superconducting state in metallic \mbox{hydrogen \cite{Ashcroft1968A}.} 

The superconducting state in hydrogenated compounds is induced by the conventional electron--phonon interaction. This fact made possible the theoretical description of the superconducting phase in ${\rm H_{2}S}$ and ${\rm H_{3}S}$ even prior to carrying out the suitable experiments \cite{Li2014A, Duan2014A}. The detailed discussion with respect to the thermodynamic properties of the superconducting state occurring in ${\rm H_{2}S}$ and ${\rm H_{3}S}$ one can find in references \cite{Durajski2015A, Duan2015A, Errea2015A, Durajski2016A, Durajski2016B, Ishikawa2016A, Errea2016A, Sano2016A, Durajski2017A, Szczesniak2018A, Kostrzewa2018A}.

In 2018, there were held the groundbreaking experiments, which confirmed the existence of the superconducting state of extremely high values of the critical temperature in the ${\rm LaH_{10}}$ compound: 
$T^{a}_{C}=215$~K for $p_{a}=150$~GPa and $T^{b}_{C}=260$~K for $p_{b}\in\left(180-200\right)$~GPa  
(and then $T^{c}_{C}\sim250$~K for $p_{c}\sim170$~GPa \cite{Drozdov2018B}). It was proved on the theoretical basis \cite{Kruglov2018A} that the results achieved by Drozdov {\it et al.} \cite{Drozdov2018A} can be related to the induction of the superconducting phase in the $R\overline{3}m$ structure ($T_{C}=206-223$~K). The experimental results reported by Somayazulu {\it et al.} \cite{Somayazulu2019A} should be related to the superconducting state induced in the $Fm\overline{3}m$ structure, where the critical temperature can potentially reach even the value of $280$~K.

From the materials science perspective, the achieved results imply that all possible actions should be taken in order to examine the hydrogen--containing materials with respect to the existence of the high--temperature superconducting state in room temperature. Attention should be paid to the importance of the discovery of the high--temperature superconducting state in ${\rm LaH_{10}}$, because La can form stable hydrogenated compounds with other metals. Such materials can exhibit so large hydrogen concentration, that they are presently taken into account as basic components of the hydrogen cells intended for vehicle drives \cite{Schlapbach2001A}.  

The purpose of this work is, firstly, to present the performed analysis of the thermodynamic properties of the superconducting state in the ${\rm LaH_{10}}$ compound. We took advantage of the phenomenological version of the Eliashberg equations, for which we fitted the value of the electron--phonon coupling constant on the basis of the experimentally found $T_{C}$ value.

Our next step consisted in examining the hydrogenated compounds of the \mbox{${\rm La_{\delta}X_{1-\delta}H_{10}}$--type} \mbox{(LaXH--type)} on the basis of the achieved results in order to find a system with even higher value of the critical temperature. Taking into account the structure of the Eliashberg function for hydrogenated compounds, with its distinctly separated parts coming from the heavy elements and from hydrogen, we assumed X to be Sc or Y, what would, in our opinion, fill the gap in the Eliashberg function occurring within the range from about $40$~meV to $100$~meV. A significant increase in the value of critical temperature should take place as a consequence.

\begin{figure}
\includegraphics[width=\columnwidth]{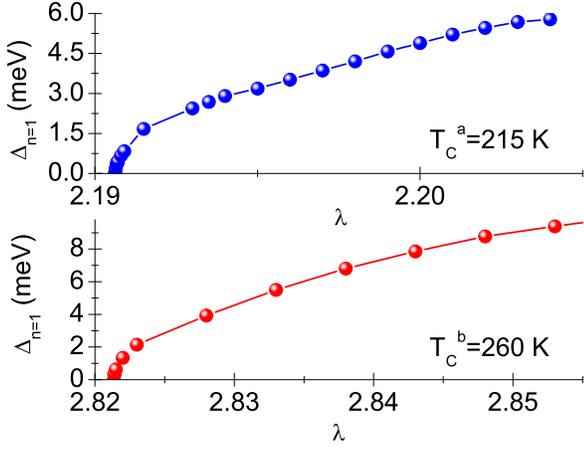}
\caption{The dependence of the maximum value of the order parameter on the electron--phonon coupling constant. 
         We consider two cases: $T^{a}_{C}=215$~K ($p_{a}=150$~GPa) and $T^{b}_{C}=260$~K ($p_{b}=190$~GPa). 
        }
\label{f1}
\end{figure}

The thermodynamic parameters of the ${\rm LaH_{10}}$ superconductor were calculated by means of Eliashberg equations on the imaginary axis \cite{Eliashberg1960A}:     
\begin{eqnarray}
\label{r01}
\Delta_{n}Z_{n}=\pi k_{B}T\sum^{M}_{m=-M}
\frac{[K\left(\omega_{n}-\omega_{m}\right)
-\mu^{\star}\left(\omega_{m}\right)]}
{\sqrt{\omega^{2}_{m}+\Delta_{m}^2}} 
{\Delta_{m}},
\end{eqnarray}
and
\begin{eqnarray}
\label{r02}
Z_{n}=1+\pi k_{B}T\sum^{M}_{m=-M}
\frac{K\left(\omega_{n}-\omega_{m}\right)}{\sqrt{\omega^{2}_{m}+
\Delta_{m}^{2}}}\frac{\omega_{m}}{\omega_{n}}Z_{m}.
\end{eqnarray}
The symbols $\Delta_{n}=\Delta\left(i\omega_n\right)$ and  $Z_{n}=Z\left(i\omega_n\right)$ denote the order parameter and the wave function renormalization factor, respectively. The quantity $\omega_n$ represents the Matsubara frequency: $\omega_{n}=\pi k_{B}T\left(2n-1\right)$, where $k_{B}$ is the Boltzmann constant. The pairing kernel is defined by: 
$K\left(\omega_n-\omega_m\right)=\lambda\frac{\Omega^2_C}{\left(\omega_n-\omega_m\right)^2+\Omega^2_C}$, where $\lambda$ denotes the electron--phonon coupling constant. We determined the value of $\lambda$ on the basis of experimental data \cite{Drozdov2018A, Somayazulu2019A} and the condition: $\left[\Delta_{n=1}\right]_{T=T_{C}}=0$. The fitting between the theory and the experimental results is presented in \fig{f1}. We obtained $\lambda_{a}=2.187$ for $p_{a}=150$~GPa and $\lambda_{b}=2.818$ for $p_{b}=190$~GPa. The symbol $\Omega_{C}$ represents the characteristic phonon frequency, its value being assumed as $\Omega_{C}=100$~meV.
\begin{figure*}
\includegraphics[width=\columnwidth]{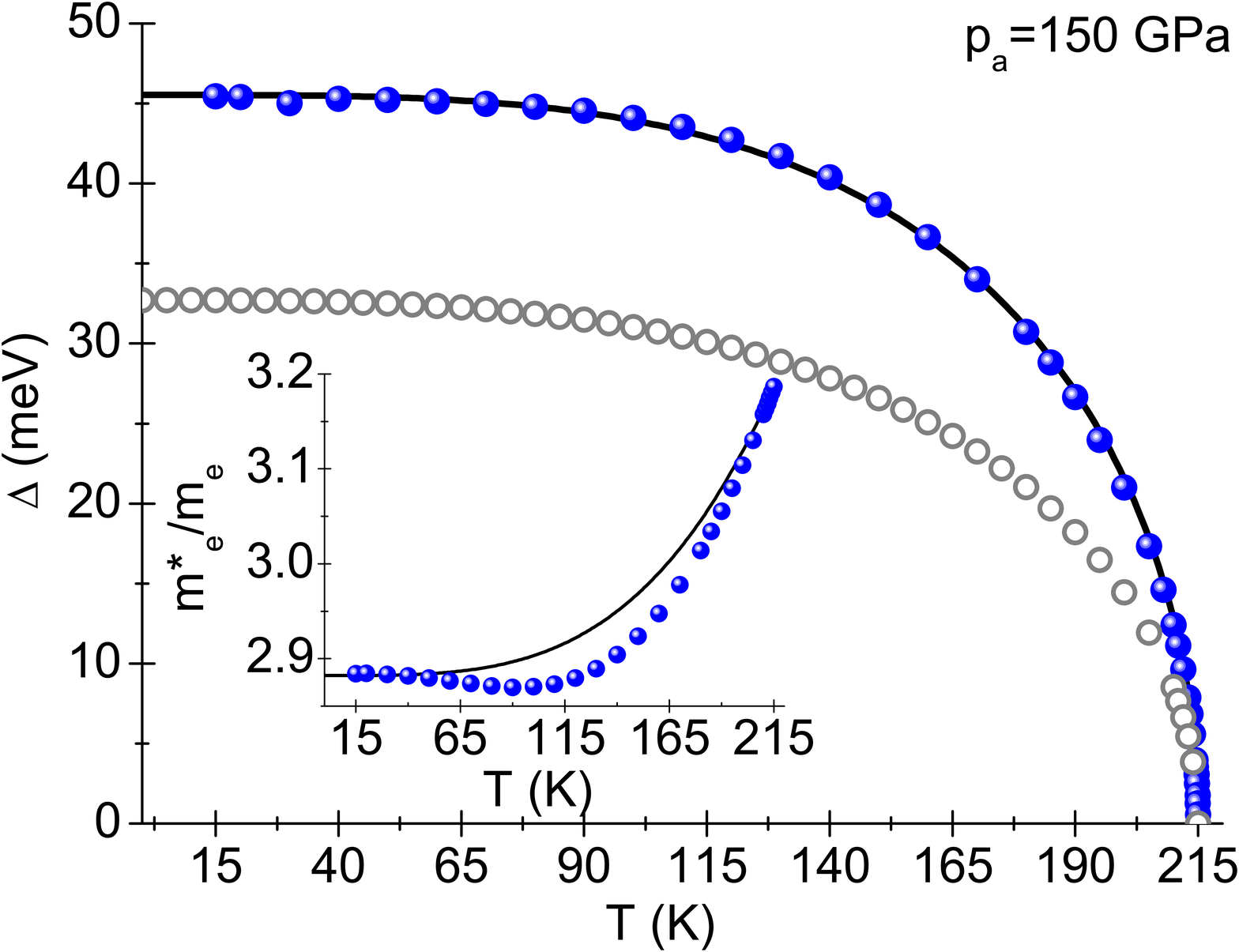}
\includegraphics[width=\columnwidth]{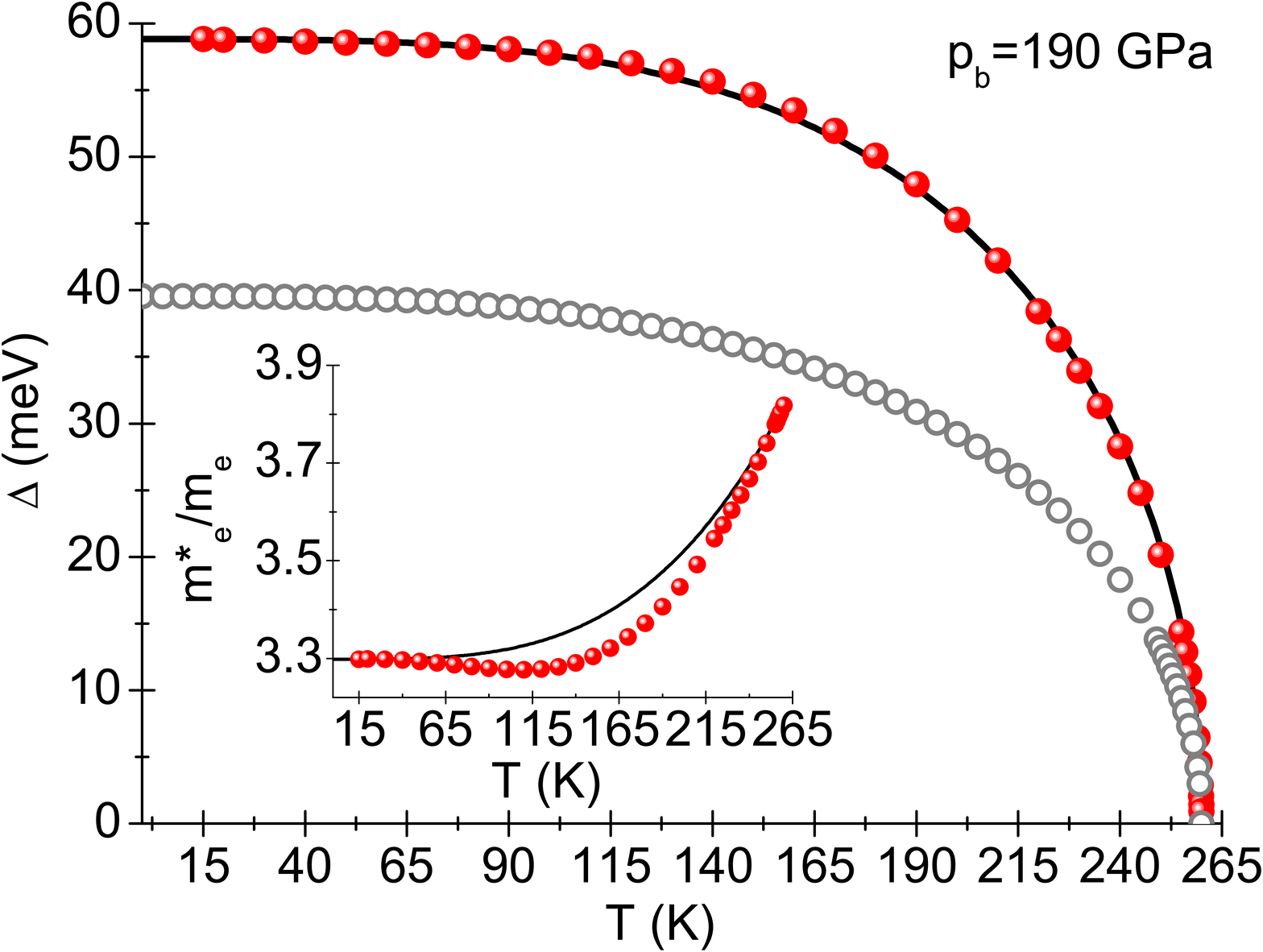}
\caption{The dependence of the order parameter on temperature.
         The insets present the influence of temperature on the value of effective electron mass to the band electron mass ratio. 
         Blue or red disks represent numerical results. Black curves were obtained from the analytical formulae: 
         $\Delta\left(T\right)=\Delta\left(T_{0}\right)\sqrt{1-\left(T\slash T_{C}\right)^{\Gamma}}$ 
         and $m_{e}^{\star}\slash m_{e}=\left[Z\left(T_{C}\right)-Z\left(T_{0}\right)\right]\left(T/T_{C}\right)^{\Gamma}
         +Z\left(T_{0}\right)$, where $Z\left(T_{C}\right)=1+\lambda$, $\Gamma_{a}=3.5$ and $\Gamma_{b}=3.4$. 
         The predictions of the BCS theory we marked with grey circles.}
\label{f2}         
\includegraphics[width=\columnwidth]{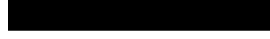}
\includegraphics[width=\columnwidth]{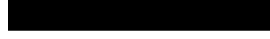}
\caption{(Bottom panels:) The difference in free energy between the superconducting and the normal state versus temperature. 
          The symbol $\rho\left(0\right)$ represents the value of the electron density of states at the Fermi surface. 
         (Top panels:) Thermodynamic critical field.
         (Insets:) The specific heat in the superconducting and the normal states.}
\label{f3}
\end{figure*}
The repulsion between electrons is modeled by the function: $\mu^{\star}\left(\omega_m\right)=\mu^{\star}\theta\left(\omega_{C}-|\omega_m|\right)$,
where $\mu^{\star}$ is the Coulomb pseudopotential ($\mu^{\star}=0.1$). The quantity $\omega_{C}$ denotes the cut--off frequency ($\omega_{C}=1$~eV). 
The Eliashberg equations were solved for the Matsubara frequency equal to $1000$. We used numerical methods presented in the previous paper \cite{Szczesniak2006A}. In the considered case, we obtained stable equation solutions for $T\geq T_{0}=15$~K.

\fig{f2} illustrates the full dependence of the order parameter on temperature. Physical values of the order parameter were calculated from the equation: $\Delta\left(T\right)={\rm Re}\left[\Delta\left(\omega=\Delta\left(T\right)\right)\right]$, 
while the function of the order parameter on the real axis ($\Delta\left(\omega\right)$) was determined using the solutions of the Eliashberg equations on the imaginary axis and the analytical continuation method described in the reference \cite{Beach2000A}. 
It can be easily seen that the order parameter curves determined within the Eliashberg formalism differ significantly from the curves resulting from the BCS theory \cite{Bardeen1957A,Bardeen1957B}. These differences arise from the very high value of the electron--phonon coupling constant of the superconductor, what is mirrored by the high value of the dimensionless $R_{\Delta}=2\Delta\left(T_{0}\right)/k_{B}T_{C}$ ratio, namely $R^{a}_{\Delta}=4.91$ and $R^{b}_{\Delta}=5.25$. Let us recall that within the BCS theory we come to the result: $\left[R_{\Delta}\right]_{\rm BCS}=3.53$, however the BCS theory approximates well the experimental results for $\lambda<0.5$.  

We plotted the temperature dependence of the effective electron mass ($m^{\star}_{e}$) to the band electron mass ($m_{e}$) ratio in the insets in \fig{f2}. The value of the $m^{\star}_{e}/m_{e}$ ratio is given with good approximation by the value of $1+\lambda$ \cite{Carbotte1990A}. 

\fig{f3} presents the results achieved for the difference in free energy between the superconducting and the normal state ($\Delta F$), the thermodynamic critical field ($H_{C}$), and the specific heat in both the superconducting ($C^{S}$) and the normal ($C^{N}$) states. The values of the considered quantities were calculated on the basis of formulae given in reference \cite{Carbotte1990A}. Deviations from the results of the BCS theory can be traced in the easiest way by determining the values of dimensionless ratios: 
$R_{H}=T_{C}C^{N}\left(T_{C}\right)/H^{2}_{C}\left(0\right)$ and $R_{C}=\Delta C\left(T_{C}\right)/C^{N}\left(T_{C}\right)$. 
For the $\rm LaH_{10}$ superconductor, we achieved the following results: $R^{a}_{H}=0.117$, $R^{b}_{H}=0.113$ and 
$R^{a}_{C}=3.51$, $R^{b}_{C}=3.75$. It is worth noticing that the BCS theory predicts $\left[R_{H}\right]_{\rm BCS}=0.168$ and $\left[R_{C}\right]_{\rm BCS}=1.43$ \cite{Bardeen1957A, Bardeen1957B, Carbotte1990A, Aguilera-Navarro1992A}. 

The subsequent last part of the paper discusses the question of induction of the superconducting state in a~group of compounds of the 
${\rm La_{\delta}X_{1-\delta}H_{10}}$--type (or LaXH--type for short). Firstly, we are going to give some criteria, which can potentially make easier the search for a material showing the required high--temperature superconducting properties. To do this, let us take into account the formula for the critical temperature valid for the BCS theory: $k_{B}T_{C}=1.13\Omega_{{\rm max}}\exp\left[-1/\rho\left(0\right)V\right]$, where $\Omega_{{\rm max}}$ denotes the Debye frequency and $V$ stands for the pairing potential value. 
It can be easily noticed that the critical temperature is the higher, the greater are the values of the electron density of states at the Fermi surface, the pairing potential, and the maximum phonon frequency. Therefore it should be supposed, even at such an early stage of analysis, that special attention is to be paid to these hydrogenated compounds, for which the respective non-hydrogenated compounds (${\rm La_{\delta}X_{1-\delta}}$) or hydrides ${\rm XH}$ exhibit the high density of electron states at the Fermi surface. Considerations given to the pairing potential at the phenomenological level do not get us very far because this quantity is calculated in a rather complicated way, usually by means of the DFT (Density Functional Theory) method.

Nevertheless, a sensible qualitative analysis can be made with respect to the influence of the atomic mass of the X element on a value of the critical temperature (since the mass of the X element determines $\Omega_{{\rm max}}$). 
In this regard, let us refer to the theoretical results obtained within the Eliashberg formalism for ${\rm H_{2}S}$ and ${\rm H_{3}S}$ superconductors \cite{Duan2014A}, \cite{Li2014A}. They prove that contributions to the Eliashberg function ($\alpha^{2}F\left(\Omega\right)$) coming from sulphur and from hydrogen are separated due to a huge difference between atomic masses of these two elements. To be precise, the electron-phonon interaction derived from sulphur is crucial in the frequency range from $0$~meV to $\Omega^{\rm S}_{\rm max}$ equal to about $70$~meV, while the contribution derived from hydrogen ($\Omega^{\rm H}_{\rm max}=220$~meV) is significant above $\sim 100$~meV. It is noteworthy that we come upon a similar situation in the case of the $\rm LaH_{10}$ compound \cite{Liu2019A}. Therefore the following factorization of the Eliashberg function for the LaXH compound can be assumed:
\begin{eqnarray}
\label{r02-B}
\alpha^{2}F\left(\Omega\right)&=&
\lambda^{\rm La}\left(\frac{\Omega}{\Omega^{\rm La}_{\rm max}}\right)^{2}\theta\left(\Omega^{\rm La}_{\rm max}-\Omega\right)\\ \nonumber
&+&
\lambda^{\rm X}\left(\frac{\Omega}{\Omega^{\rm X}_{\rm max}}\right)^{2}\theta\left(\Omega^{\rm X}_{\rm max}-\Omega\right)\\ \nonumber
&+&
\lambda^{\rm H}\left(\frac{\Omega}{\Omega^{\rm H}_{\rm max}}\right)^{2}\theta\left(\Omega^{\rm H}_{\rm max}-\Omega\right),
\end{eqnarray}
where $\lambda^{\rm La}$, $\lambda^{\rm X}$, and $\lambda^{\rm H}$ are the contributions to the electron--phonon coupling constant derived from both metals (La, X) and hydrogen, respectively. Similarly, the symbols $\Omega^{\rm La}_{\rm max}$, $\Omega^{\rm X}_{\rm max}$, and $\Omega^{\rm H}_{\rm max}$ represent the respective maximum phonon frequencies. The value of the critical temperature can be assessed from the generalised formula of the BCS theory \cite{Durajski2015A}:   
\begin{equation}
\label{r03-B}
k_{B}T_{C}=f_{1}f_{2}\frac{\omega_{{\rm ln}}}{1.27}\exp\left[\frac{-1.14\left(1+\lambda\right)}{\lambda-(1 + 0.163 \lambda)\mu^{\star}}\right],
\end{equation}
while the symbols appearing in \eq{r03-B} are defined in \mbox{\tab{t1}}.
\begin{table}
\caption{\label{t1}
The quantities: $\lambda$, $\omega_{{\rm ln}}$ (logarithmic phonon frequency), 
$\omega_{2}$ (second moment of the normalized weight function), 
$f_{1}$ (strong--coupling correction function), and 
$f_{2}$ (shape correction function).}
\begin{ruledtabular}
\begin{tabular}{c}
Quantity \\
\hline
 \\
${\lambda= 2\int^{+\infty}_0 d\Omega \frac{\alpha^2\left(\Omega\right)F\left(\Omega\right)}{\Omega}}$,\\
\\
$\omega_{{\rm ln}}=\exp\left[\frac{2}{\lambda}\int^{+\infty}_{0}d\Omega\frac{\alpha^{2}F\left(\Omega\right)}{\Omega}\ln\left(\Omega\right)\right]$,\\
 \\ 
$\omega_{2}=\frac{2}{\lambda}\int^{+\infty}_{0}d\Omega\alpha^{2}F\left(\Omega\right)\Omega$,\\
 \\
$f_{1}=\left[1+\left(\frac{\lambda}{\Lambda_{1}}\right)^{\frac{3}{2}}\right]^{\frac{1}{3}}$, \hspace{1mm}
$f_{2}=1+\frac{\left(\frac{\sqrt{\omega_{2}}}{\omega_{\rm{ln}}}-1\right)\lambda^{2}}{\lambda^{2}+\Lambda^{2}_{2}}$,\\
 \\
$\Lambda_{1}=2.4-0.14\mu^{\star}$,\\
 \\
$\Lambda_{2}=\left(0.1+9\mu^{\star}\right)\left(\sqrt{\omega_{2}}/\omega_{\ln}\right)$.\\
 \\
\end{tabular}
\end{ruledtabular}
\end{table}
\begin{figure*}
\includegraphics[width=\columnwidth]{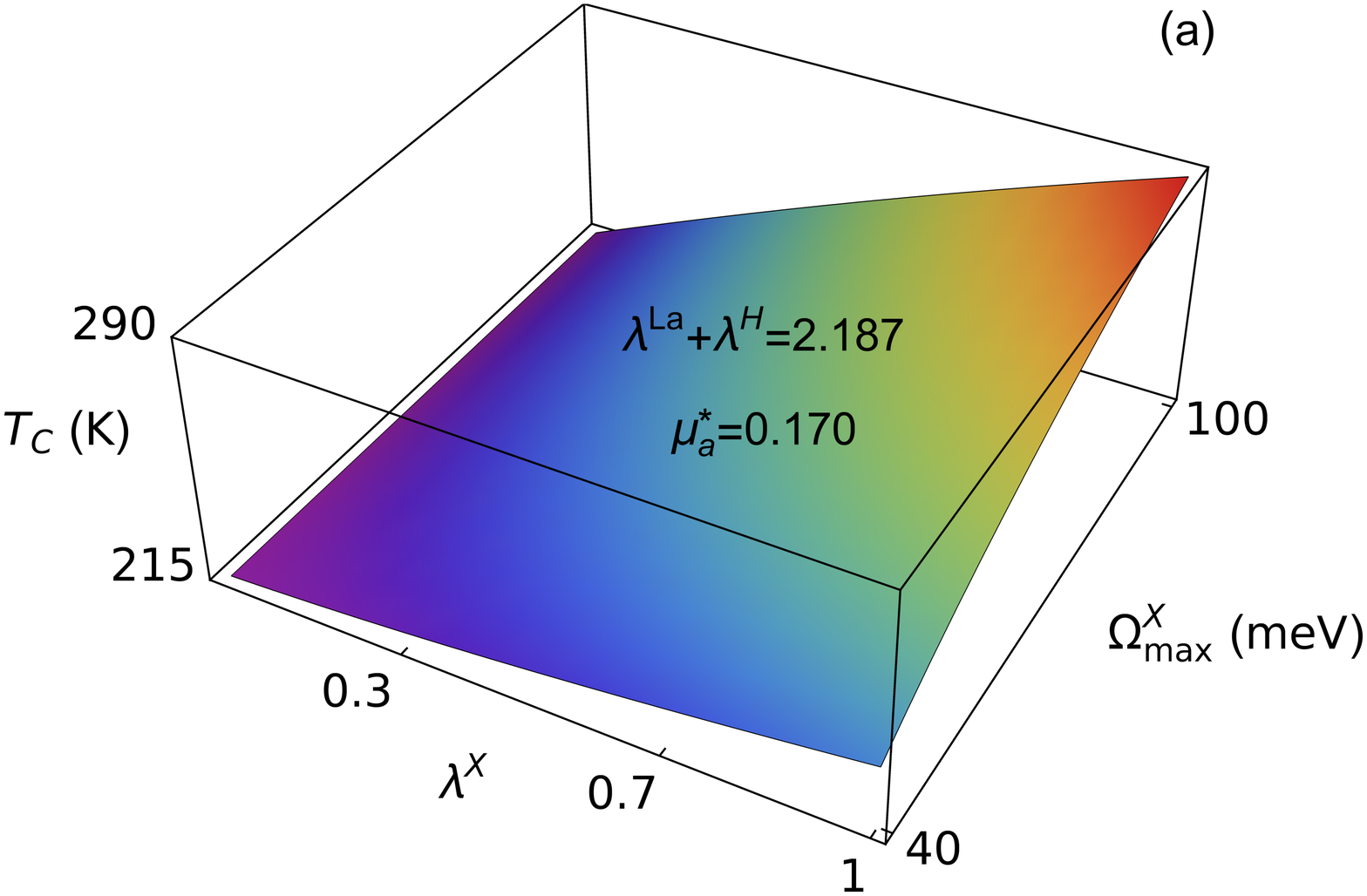}
\includegraphics[width=\columnwidth]{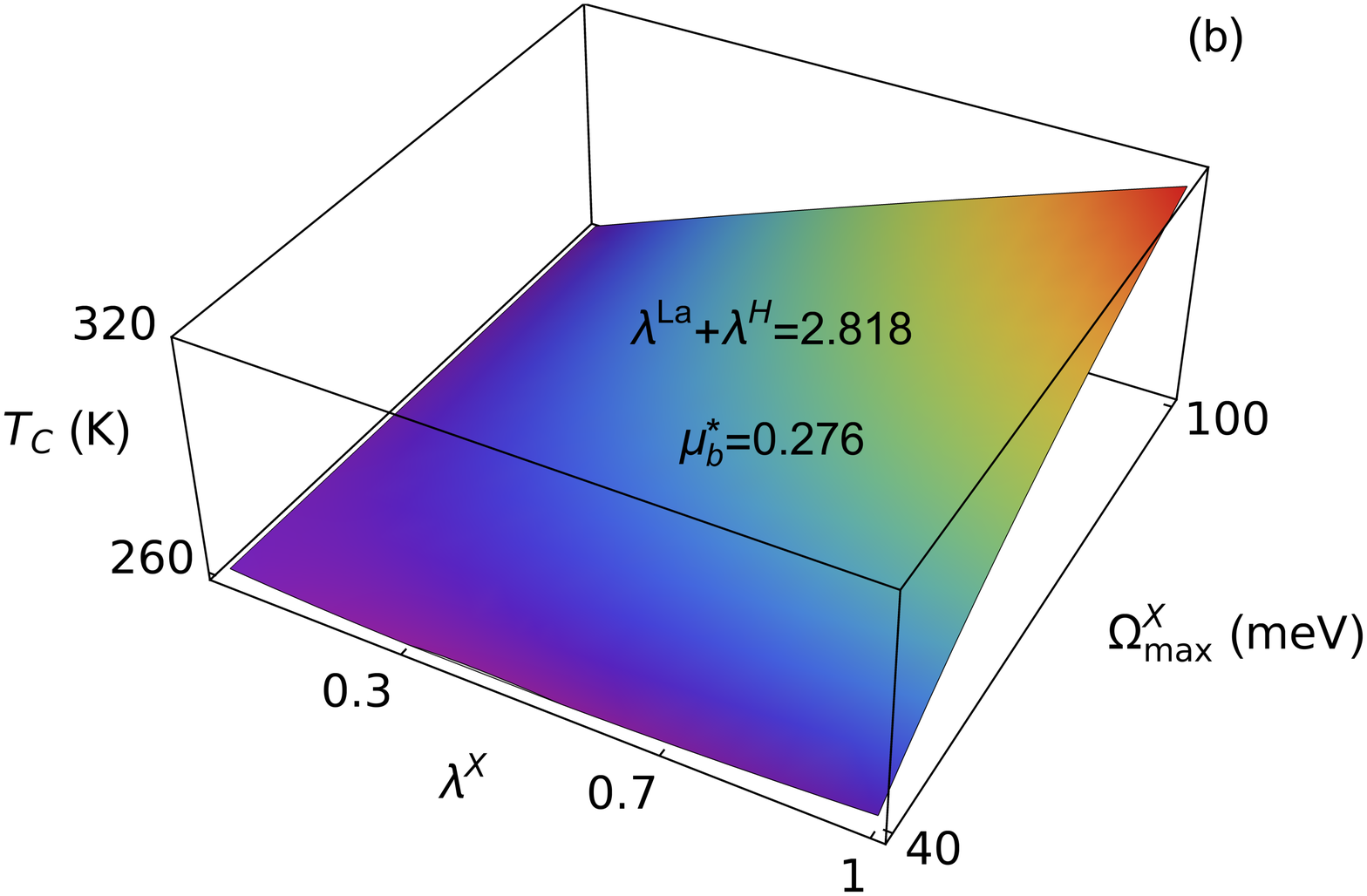}
\caption{The dependence of the critical temperature on $\lambda^{\rm X}$ and $\Omega^{\rm X}_{\rm max}$.
         Figure (a) presents the results for $\lambda^{\rm La}$+ $\lambda^{\rm H}$=2.187 and $\mu^{\star}_{a}=0.170$. 
         Figure (b) is plotted for $\lambda^{\rm La}$+ $\lambda^{\rm H}=2.818$ and $\mu^{\star}_{b}=0.276$. 
         It was assumed that $\Omega^{\rm La}_{\rm max}=40$~meV and $\Omega^{\rm H}_{\rm max}=290$~meV for both cases.}
\label{f4}
\end{figure*}
\begin{figure}
\includegraphics[width=\columnwidth]{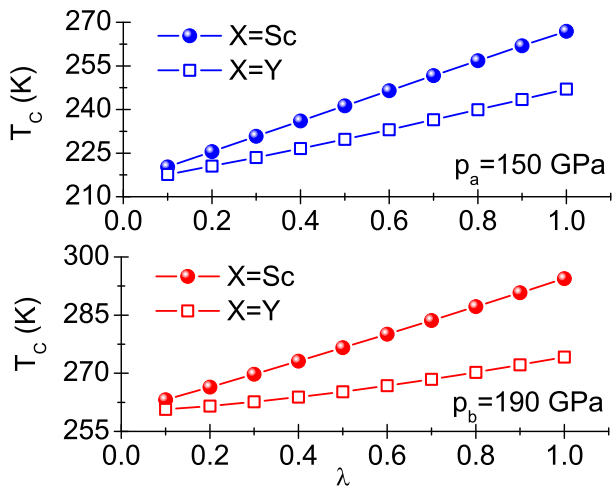}
\caption{The expected range of critical temperature values for the LaScH and the LaYH compounds.}
\label{f5}
\end{figure}

Let us calculate explicitly the relevant quantities:
\begin{equation}
\label{r04-B}
\lambda=\lambda^{\rm La}+\lambda^{\rm X}+\lambda^{\rm H},
\end{equation}
\begin{eqnarray}
\label{r05-B}
\omega_{\rm ln}&=&
{\rm Exp}\left[\frac{\lambda^{\rm La}}{\lambda^{\rm La}+\lambda^{\rm X}+\lambda^{\rm H}}\left(\ln\left(\Omega^{\rm La}_{\rm max}\right)-\frac{1}{2}\right)\right]\\ \nonumber
&\times&
{\rm Exp}\left[\frac{\lambda^{\rm X}}{\lambda^{\rm La}+\lambda^{\rm X}+\lambda^{\rm H}}\left(\ln\left(\Omega^{\rm X}_{\rm max}\right)-\frac{1}{2}\right)\right]\\ \nonumber
&\times&
{\rm Exp}\left[\frac{\lambda^{\rm H}}{\lambda^{\rm La}+\lambda^{\rm X}+\lambda^{\rm H}}\left(\ln\left(\Omega^{\rm H}_{\rm max}\right)-\frac{1}{2}\right)\right],
\end{eqnarray}
and
\begin{eqnarray}
\label{r06-B}
\omega_{2}&=&\frac{\lambda^{\rm La}}{\lambda^{\rm La}+\lambda^{\rm X}+\lambda^{\rm H}}\frac{\left(\Omega^{\rm La}_{\rm max}\right)^{2}}{2}
\\ \nonumber
&+&
\frac{\lambda^{\rm X}}{\lambda^{\rm La}+\lambda^{\rm X}+\lambda^{\rm H}}\frac{\left(\Omega^{\rm X}_{\rm max}\right)^{2}}{2}
\\ \nonumber
&+&
\frac{\lambda^{\rm H}}{\lambda^{\rm La}+\lambda^{\rm X}+\lambda^{\rm H}}\frac{\left(\Omega^{\rm H}_{\rm max}\right)^{2}}{2}.
\end{eqnarray}

We are going to consider the case $\Omega^{\rm La}_{\rm max}\sim 40\hspace{1mm}{\rm meV}<\Omega^{\rm X}_{\rm max}<100\hspace{1mm}{\rm meV}$. It means that we are interested in such an X element, the contribution of which to the Eliashberg function fills the gap between contributions coming from lanthanum and from hydrogen. It can be assumed that $0<\lambda^{\rm X}<1$, while keeping in mind that $\lambda^{\rm La}=0.68$ \cite{Chen2019A}. Additionally, the previous calculations discussed in the work allow to write that $\lambda^{\rm La}+\lambda^{\rm H}$ is equal to $\lambda_{a}=2.187$ for $p_{a}=150$~GPa or to $\lambda_{b}=2.818$ for $p_{b}=190$~GPa. The quantity $\mu^{\star}$ occurring in the \eq{r03-B} serves now as the fitting parameter. One should remember that the formula for the critical temperature given by the \eq{r03-B} was derived with the use of significant simplyfing assumptions (the value of the cut--off frequency is neglected, as well as the retardation effects modelled by the Matsubara frequency). Therefore the value of the Coulomb pseudopotential determined from the full Eliashberg equations usually differs from the value of $\mu^{\star}$ calculated analytically. The experimental data for the ${\rm LaH_{10}}$ superconductor can be reproduced using \eq{r03-B} and assuming that $\mu^{\star}_{a}=0.170$ and $\mu^{\star}_{b}=0.276$.

The achieved results are presented in \fig{f4}. It is evident that taking into consideration the additional X element, which enriches the ${\rm LaH_{10}}$ composition, leads to a large increase in the critical temperature value. The estimated upper limit of the $T^{a}_{C}$ value is equal to $288$~K for $p_{a}=150$~GPa, while for $p_{b}=190$~GPa we obtain $T^{b}_{C}=315$~K. Therefore the superconducting state can potentially exists in room temperature for both cases.

Now, let us take into account elements with the identical electron configuration at the valence shell as lanthanum, but lighter than lanthanum: scandium and yttrium, both being selected as X. Attention should be paid to the fact that the electron configuration of X, identical as in lanthanum, should minimize such changes in properties of the obtained compound which could result from changes in both the electron dispersion relation and the matrix elements of the electron--phonon interaction. Applying the formula: 
$\Omega^{\rm X}_{\rm max}/\Omega^{\rm La}_{\rm max}\sim \sqrt{M_{\rm La}/M_{\rm X}}$ we get $\Omega^{\rm Sc}_{\rm max}\sim 70$~meV and 
$\Omega^{\rm Y}_{\rm max}\sim 50$~meV ($M_{\rm La}$ and $M_{\rm X}$ denote atomic mass of lanthanum and the element X, i.e. Sc or Y, respectively). 

\fig{f5} presents the expected range of the critical temperature values for the LaScH and the LaYH compounds. We took into account two pressure values: $p_{a}=150$~GPa and $p_{b}=190$~GPa. For LaScH we got: $T^{a}_{C}\in\left<220,267\right>$~K and $T^{b}_{C}\in\left<263,294\right>$~K, while the results for LaYH are as follows: 
$T^{a}_{C}\in\left<218,247\right>$~K and $T^{b}_{C}\in\left<261,274\right>$~K. Apparently, the significant increase in the critical temperature value should be observed in both cases. The effect of growth in the value of the critical temperature results from filling the gap in the Eliashberg function between the contributions coming from La and H, as was already stated above.

To summarize, the experimental results obtained for the ${\rm LaH_{10}}$ compound get us much closer to the purpose of obtaining the superconducting state at room temperature. The huge difference between atomic masses of lanthanum and hydrogen results in the characteristic structure of the Eliashberg function modeling the electron--phonon interaction in the considered compound, with distinctly separated parts proceeded either from lanthanum or from hydrogen. The proper selection of the additional element (X) in the LaXH compound is expected to fill the \mbox{'empty' range} of the Eliashberg function between the parts coming from La and H. In our opinion, good candidates are scandium and yttrium. These elements have the electron configuration at the valence shell exactly the same as lanthanum, and yet they are considerably lighter. Our numerical calculations suggest the possible growth in the critical temperature of the LaScH compound equal to about $52$~K ($150$~GPa) or to about $79$~K ($190$~GPa) as compared to the $T_{C}$ value for the ${\rm LaH_{10}}$ compound. As far as the LaYH compound is concerned, the pertinent increase in $T_{C}$ value can reach about $32$~K for $150$~GPa or about $59$~K for $190$~GPa. Our results can be a starting point for the advanced DFT calculations or perhaps provide inspiration for carrying out the appropriate experimental measurements.

One needs to recognise that the exact quantitative analysis of the problem discussed in our work would require to be carried out using the Eliashberg equations including the anharmonicity of the phonon system and the non-linear terms of the electron--phonon--phonon interaction, especially for such high values of the critical temperature as are observed for ${\rm LaH_{10}}$. Presently we work upon the derivation of suitable equations.  

\bibliography{Bibliography}

\begin{thebibliography}{31}
\expandafter\ifx\csname natexlab\endcsname\relax\def\natexlab#1{#1}\fi
\expandafter\ifx\csname bibnamefont\endcsname\relax
  \def\bibnamefont#1{#1}\fi
\expandafter\ifx\csname bibfnamefont\endcsname\relax
  \def\bibfnamefont#1{#1}\fi
\expandafter\ifx\csname citenamefont\endcsname\relax
  \def\citenamefont#1{#1}\fi
\expandafter\ifx\csname url\endcsname\relax
  \def\url#1{\texttt{#1}}\fi
\expandafter\ifx\csname urlprefix\endcsname\relax\def\urlprefix{URL }\fi
\providecommand{\bibinfo}[2]{#2}
\providecommand{\eprint}[2][]{\url{#2}}

\bibitem[{\citenamefont{Drozdov et~al.}(2014)\citenamefont{Drozdov, Eremets,
  and Troyan}}]{Drozdov2014A}
\bibinfo{author}{\bibfnamefont{A.~P.} \bibnamefont{Drozdov}},
  \bibinfo{author}{\bibfnamefont{M.~I.} \bibnamefont{Eremets}},
  \bibnamefont{and} \bibinfo{author}{\bibfnamefont{I.~A.}
  \bibnamefont{Troyan}}, \bibinfo{journal}{arXiv: 1412.0460}
  (\bibinfo{year}{2014}).

\bibitem[{\citenamefont{Drozdov et~al.}(2015)\citenamefont{Drozdov, Eremets,
  Troyan, Ksenofontov, and Shylin}}]{Drozdov2015A}
\bibinfo{author}{\bibfnamefont{A.~P.} \bibnamefont{Drozdov}},
  \bibinfo{author}{\bibfnamefont{M.~I.} \bibnamefont{Eremets}},
  \bibinfo{author}{\bibfnamefont{I.~A.} \bibnamefont{Troyan}},
  \bibinfo{author}{\bibfnamefont{V.}~\bibnamefont{Ksenofontov}},
  \bibnamefont{and} \bibinfo{author}{\bibfnamefont{S.~I.}
  \bibnamefont{Shylin}}, \bibinfo{journal}{Nature}
  \textbf{\bibinfo{volume}{525}}, \bibinfo{pages}{73} (\bibinfo{year}{2015}).

\bibitem[{\citenamefont{Ashcroft}(2004)}]{Ashcroft2004A}
\bibinfo{author}{\bibfnamefont{N.~W.} \bibnamefont{Ashcroft}},
  \bibinfo{journal}{Physical Review Letters} \textbf{\bibinfo{volume}{92}},
  \bibinfo{pages}{187002} (\bibinfo{year}{2004}).

\bibitem[{\citenamefont{Ashcroft}(1968)}]{Ashcroft1968A}
\bibinfo{author}{\bibfnamefont{N.~W.} \bibnamefont{Ashcroft}},
  \bibinfo{journal}{Physical Review Letters} \textbf{\bibinfo{volume}{21}},
  \bibinfo{pages}{1748} (\bibinfo{year}{1968}).

\bibitem[{\citenamefont{Li et~al.}(2014)\citenamefont{Li, Hao, Liu, Li, and
  Ma}}]{Li2014A}
\bibinfo{author}{\bibfnamefont{Y.}~\bibnamefont{Li}},
  \bibinfo{author}{\bibfnamefont{J.}~\bibnamefont{Hao}},
  \bibinfo{author}{\bibfnamefont{H.}~\bibnamefont{Liu}},
  \bibinfo{author}{\bibfnamefont{Y.}~\bibnamefont{Li}}, \bibnamefont{and}
  \bibinfo{author}{\bibfnamefont{Y.}~\bibnamefont{Ma}}, \bibinfo{journal}{The
  Journal of Chemical Physics} \textbf{\bibinfo{volume}{140}},
  \bibinfo{pages}{174712} (\bibinfo{year}{2014}).

\bibitem[{\citenamefont{Duan et~al.}(2014)\citenamefont{Duan, Liu, Tian, Li,
  Huang, Zhao, Yu, Liu, Tian, and Cui}}]{Duan2014A}
\bibinfo{author}{\bibfnamefont{D.}~\bibnamefont{Duan}},
  \bibinfo{author}{\bibfnamefont{Y.}~\bibnamefont{Liu}},
  \bibinfo{author}{\bibfnamefont{F.}~\bibnamefont{Tian}},
  \bibinfo{author}{\bibfnamefont{D.}~\bibnamefont{Li}},
  \bibinfo{author}{\bibfnamefont{X.}~\bibnamefont{Huang}},
  \bibinfo{author}{\bibfnamefont{Z.}~\bibnamefont{Zhao}},
  \bibinfo{author}{\bibfnamefont{H.}~\bibnamefont{Yu}},
  \bibinfo{author}{\bibfnamefont{B.}~\bibnamefont{Liu}},
  \bibinfo{author}{\bibfnamefont{W.}~\bibnamefont{Tian}}, \bibnamefont{and}
  \bibinfo{author}{\bibfnamefont{T.}~\bibnamefont{Cui}},
  \bibinfo{journal}{Scientific Reports} \textbf{\bibinfo{volume}{4}},
  \bibinfo{pages}{6968} (\bibinfo{year}{2014}).

\bibitem[{\citenamefont{Durajski et~al.}(2015)\citenamefont{Durajski,
  Szcz{\c{e}}{\'s}niak, and Li}}]{Durajski2015A}
\bibinfo{author}{\bibfnamefont{A.~P.} \bibnamefont{Durajski}},
  \bibinfo{author}{\bibfnamefont{R.}~\bibnamefont{Szcz{\c{e}}{\'s}niak}},
  \bibnamefont{and} \bibinfo{author}{\bibfnamefont{Y.}~\bibnamefont{Li}},
  \bibinfo{journal}{Physica C} \textbf{\bibinfo{volume}{515}},
  \bibinfo{pages}{1} (\bibinfo{year}{2015}).

\bibitem[{\citenamefont{Duan et~al.}(2015)\citenamefont{Duan, Huang, Tian, Li,
  Yu, Liu, Ma, Liu, and Cui}}]{Duan2015A}
\bibinfo{author}{\bibfnamefont{D.}~\bibnamefont{Duan}},
  \bibinfo{author}{\bibfnamefont{X.}~\bibnamefont{Huang}},
  \bibinfo{author}{\bibfnamefont{F.}~\bibnamefont{Tian}},
  \bibinfo{author}{\bibfnamefont{D.}~\bibnamefont{Li}},
  \bibinfo{author}{\bibfnamefont{H.}~\bibnamefont{Yu}},
  \bibinfo{author}{\bibfnamefont{Y.}~\bibnamefont{Liu}},
  \bibinfo{author}{\bibfnamefont{Y.}~\bibnamefont{Ma}},
  \bibinfo{author}{\bibfnamefont{B.}~\bibnamefont{Liu}}, \bibnamefont{and}
  \bibinfo{author}{\bibfnamefont{T.}~\bibnamefont{Cui}},
  \bibinfo{journal}{Physical Review B} \textbf{\bibinfo{volume}{91}},
  \bibinfo{pages}{180502(R)} (\bibinfo{year}{2015}).

\bibitem[{\citenamefont{Errea et~al.}(2015)\citenamefont{Errea, Calandra,
  Pickard, Richard, Needs, Li, Liu, Zhang, Ma, and Mauri}}]{Errea2015A}
\bibinfo{author}{\bibfnamefont{I.}~\bibnamefont{Errea}},
  \bibinfo{author}{\bibfnamefont{M.}~\bibnamefont{Calandra}},
  \bibinfo{author}{\bibfnamefont{C.~J.} \bibnamefont{Pickard}},
  \bibinfo{author}{\bibfnamefont{J.~N.} \bibnamefont{Richard}},
  \bibinfo{author}{\bibfnamefont{J.}~\bibnamefont{Needs}},
  \bibinfo{author}{\bibfnamefont{Y.}~\bibnamefont{Li}},
  \bibinfo{author}{\bibfnamefont{H.}~\bibnamefont{Liu}},
  \bibinfo{author}{\bibfnamefont{Y.}~\bibnamefont{Zhang}},
  \bibinfo{author}{\bibfnamefont{Y.}~\bibnamefont{Ma}}, \bibnamefont{and}
  \bibinfo{author}{\bibfnamefont{F.}~\bibnamefont{Mauri}},
  \bibinfo{journal}{Physical Review Letters} \textbf{\bibinfo{volume}{114}},
  \bibinfo{pages}{157004} (\bibinfo{year}{2015}).

\bibitem[{\citenamefont{Durajski}(2016)}]{Durajski2016A}
\bibinfo{author}{\bibfnamefont{A.~P.} \bibnamefont{Durajski}},
  \bibinfo{journal}{Scientific Reports} \textbf{\bibinfo{volume}{6}},
  \bibinfo{pages}{38570} (\bibinfo{year}{2016}).

\bibitem[{\citenamefont{Durajski et~al.}(2016)\citenamefont{Durajski,
  Szcz{\c{e}}{\'s}niak, and Pietronero}}]{Durajski2016B}
\bibinfo{author}{\bibfnamefont{A.~P.} \bibnamefont{Durajski}},
  \bibinfo{author}{\bibfnamefont{R.}~\bibnamefont{Szcz{\c{e}}{\'s}niak}},
  \bibnamefont{and}
  \bibinfo{author}{\bibfnamefont{L.}~\bibnamefont{Pietronero}},
  \bibinfo{journal}{Annalen der Physik} \textbf{\bibinfo{volume}{528}},
  \bibinfo{pages}{358} (\bibinfo{year}{2016}).

\bibitem[{\citenamefont{Ishikawa et~al.}(2016)\citenamefont{Ishikawa,
  Nakanishi, Shimizu, Katayama-Yoshida, Oda, and Suzuki}}]{Ishikawa2016A}
\bibinfo{author}{\bibfnamefont{T.}~\bibnamefont{Ishikawa}},
  \bibinfo{author}{\bibfnamefont{A.}~\bibnamefont{Nakanishi}},
  \bibinfo{author}{\bibfnamefont{K.}~\bibnamefont{Shimizu}},
  \bibinfo{author}{\bibfnamefont{H.}~\bibnamefont{Katayama-Yoshida}},
  \bibinfo{author}{\bibfnamefont{T.}~\bibnamefont{Oda}}, \bibnamefont{and}
  \bibinfo{author}{\bibfnamefont{N.}~\bibnamefont{Suzuki}},
  \bibinfo{journal}{Scientific Reports} \textbf{\bibinfo{volume}{6}},
  \bibinfo{pages}{23160} (\bibinfo{year}{2016}).

\bibitem[{\citenamefont{Errea et~al.}(2016)\citenamefont{Errea, Calandra,
  Pickard, Nelson, Needs, Li, Liu, Zhang, Ma, and Mauri}}]{Errea2016A}
\bibinfo{author}{\bibfnamefont{I.}~\bibnamefont{Errea}},
  \bibinfo{author}{\bibfnamefont{M.}~\bibnamefont{Calandra}},
  \bibinfo{author}{\bibfnamefont{C.~J.} \bibnamefont{Pickard}},
  \bibinfo{author}{\bibfnamefont{J.~R.} \bibnamefont{Nelson}},
  \bibinfo{author}{\bibfnamefont{R.~J.} \bibnamefont{Needs}},
  \bibinfo{author}{\bibfnamefont{Y.}~\bibnamefont{Li}},
  \bibinfo{author}{\bibfnamefont{H.}~\bibnamefont{Liu}},
  \bibinfo{author}{\bibfnamefont{Y.}~\bibnamefont{Zhang}},
  \bibinfo{author}{\bibfnamefont{Y.}~\bibnamefont{Ma}}, \bibnamefont{and}
  \bibinfo{author}{\bibfnamefont{F.}~\bibnamefont{Mauri}},
  \bibinfo{journal}{Nature} \textbf{\bibinfo{volume}{532}}, \bibinfo{pages}{81}
  (\bibinfo{year}{2016}).

\bibitem[{\citenamefont{Sano et~al.}(2016)\citenamefont{Sano, Koretsune,
  Tadano, Akashi, and Arita}}]{Sano2016A}
\bibinfo{author}{\bibfnamefont{W.}~\bibnamefont{Sano}},
  \bibinfo{author}{\bibfnamefont{T.}~\bibnamefont{Koretsune}},
  \bibinfo{author}{\bibfnamefont{T.}~\bibnamefont{Tadano}},
  \bibinfo{author}{\bibfnamefont{R.}~\bibnamefont{Akashi}}, \bibnamefont{and}
  \bibinfo{author}{\bibfnamefont{R.}~\bibnamefont{Arita}},
  \bibinfo{journal}{Physical Review B} \textbf{\bibinfo{volume}{93}},
  \bibinfo{pages}{094525} (\bibinfo{year}{2016}).

\bibitem[{\citenamefont{Durajski and
  Szcz{\c{e}}{\'s}niak}(2017)}]{Durajski2017A}
\bibinfo{author}{\bibfnamefont{A.~P.} \bibnamefont{Durajski}} \bibnamefont{and}
  \bibinfo{author}{\bibfnamefont{R.}~\bibnamefont{Szcz{\c{e}}{\'s}niak}},
  \bibinfo{journal}{Scientific Reports} \textbf{\bibinfo{volume}{7}},
  \bibinfo{pages}{4473} (\bibinfo{year}{2017}).

\bibitem[{\citenamefont{Szcz{\c{e}}{\'s}niak and
  Durajski}(2018)}]{Szczesniak2018A}
\bibinfo{author}{\bibfnamefont{R.}~\bibnamefont{Szcz{\c{e}}{\'s}niak}}
  \bibnamefont{and} \bibinfo{author}{\bibfnamefont{A.~P.}
  \bibnamefont{Durajski}}, \bibinfo{journal}{Scientific Reports}
  \textbf{\bibinfo{volume}{8}}, \bibinfo{pages}{6037} (\bibinfo{year}{2018}).

\bibitem[{\citenamefont{Kostrzewa et~al.}(2018)\citenamefont{Kostrzewa,
  Szcz{\c{e}}{\'s}niak, Kalaga, and Wrona}}]{Kostrzewa2018A}
\bibinfo{author}{\bibfnamefont{M.}~\bibnamefont{Kostrzewa}},
  \bibinfo{author}{\bibfnamefont{R.}~\bibnamefont{Szcz{\c{e}}{\'s}niak}},
  \bibinfo{author}{\bibfnamefont{J.~K.} \bibnamefont{Kalaga}},
  \bibnamefont{and} \bibinfo{author}{\bibfnamefont{I.~A.} \bibnamefont{Wrona}},
  \bibinfo{journal}{Scientific Reports} \textbf{\bibinfo{volume}{8}},
  \bibinfo{pages}{11957} (\bibinfo{year}{2018}).

\bibitem[{\citenamefont{Drozdov
  et~al.}(2018{\natexlab{a}})\citenamefont{Drozdov, Minkov, Besedin, Kong,
  Kuzovnikov, D., Mozaffari, Balicas, Balakirev, Graf et~al.}}]{Drozdov2018B}
\bibinfo{author}{\bibfnamefont{A.~P.} \bibnamefont{Drozdov}},
  \bibinfo{author}{\bibfnamefont{V.~S.} \bibnamefont{Minkov}},
  \bibinfo{author}{\bibfnamefont{S.~P.} \bibnamefont{Besedin}},
  \bibinfo{author}{\bibfnamefont{P.~P.} \bibnamefont{Kong}},
  \bibinfo{author}{\bibfnamefont{M.~A.} \bibnamefont{Kuzovnikov}},
  \bibinfo{author}{\bibnamefont{D.}},
  \bibinfo{author}{\bibfnamefont{S.}~\bibnamefont{Mozaffari}},
  \bibinfo{author}{\bibfnamefont{L.}~\bibnamefont{Balicas}},
  \bibinfo{author}{\bibfnamefont{F.}~\bibnamefont{Balakirev}},
  \bibinfo{author}{\bibfnamefont{D.}~\bibnamefont{Graf}}, \bibnamefont{et~al.},
  \bibinfo{journal}{arXiv:1812.01561}  (\bibinfo{year}{2018}{\natexlab{a}}).

\bibitem[{\citenamefont{Kruglov et~al.}(2018)\citenamefont{Kruglov, Semenok,
  Szcz{\c{e}}{\'s}niak, Esfahani, Kvashnin, and Organov}}]{Kruglov2018A}
\bibinfo{author}{\bibfnamefont{I.~A.} \bibnamefont{Kruglov}},
  \bibinfo{author}{\bibfnamefont{D.~V.} \bibnamefont{Semenok}},
  \bibinfo{author}{\bibfnamefont{R.}~\bibnamefont{Szcz{\c{e}}{\'s}niak}},
  \bibinfo{author}{\bibfnamefont{M.~M.~D.} \bibnamefont{Esfahani}},
  \bibinfo{author}{\bibfnamefont{A.~G.} \bibnamefont{Kvashnin}},
  \bibnamefont{and} \bibinfo{author}{\bibfnamefont{A.~R.}
  \bibnamefont{Organov}}, \bibinfo{journal}{arXiv:1810.0111}
  (\bibinfo{year}{2018}).

\bibitem[{\citenamefont{Drozdov
  et~al.}(2018{\natexlab{b}})\citenamefont{Drozdov, Minkov, Besedin, Kong,
  Kuzovnikov, Knyazev, and Eremets}}]{Drozdov2018A}
\bibinfo{author}{\bibfnamefont{A.~P.} \bibnamefont{Drozdov}},
  \bibinfo{author}{\bibfnamefont{V.~S.} \bibnamefont{Minkov}},
  \bibinfo{author}{\bibfnamefont{S.~P.} \bibnamefont{Besedin}},
  \bibinfo{author}{\bibfnamefont{P.~P.} \bibnamefont{Kong}},
  \bibinfo{author}{\bibfnamefont{M.~A.} \bibnamefont{Kuzovnikov}},
  \bibinfo{author}{\bibfnamefont{D.~A.} \bibnamefont{Knyazev}},
  \bibnamefont{and} \bibinfo{author}{\bibfnamefont{M.~I.}
  \bibnamefont{Eremets}}, \bibinfo{journal}{arXiv:1808.07039}
  (\bibinfo{year}{2018}{\natexlab{b}}).

\bibitem[{\citenamefont{Somayazulu et~al.}(2019)\citenamefont{Somayazulu,
  Ahart, Mishra, Geballe, Baldini, Meng, Struzhkin, and
  Hemley}}]{Somayazulu2019A}
\bibinfo{author}{\bibfnamefont{M.}~\bibnamefont{Somayazulu}},
  \bibinfo{author}{\bibfnamefont{M.}~\bibnamefont{Ahart}},
  \bibinfo{author}{\bibfnamefont{A.~K.} \bibnamefont{Mishra}},
  \bibinfo{author}{\bibfnamefont{Z.~M.} \bibnamefont{Geballe}},
  \bibinfo{author}{\bibfnamefont{M.}~\bibnamefont{Baldini}},
  \bibinfo{author}{\bibfnamefont{Y.}~\bibnamefont{Meng}},
  \bibinfo{author}{\bibfnamefont{V.~V.} \bibnamefont{Struzhkin}},
  \bibnamefont{and} \bibinfo{author}{\bibfnamefont{R.~J.}
  \bibnamefont{Hemley}}, \bibinfo{journal}{Physical Review Letters}
  \textbf{\bibinfo{volume}{122}}, \bibinfo{pages}{027001}
  (\bibinfo{year}{2019}).

\bibitem[{\citenamefont{L.Schlapbach and Z{\"u}ttel}(2001)}]{Schlapbach2001A}
\bibinfo{author}{\bibnamefont{L.Schlapbach}} \bibnamefont{and}
  \bibinfo{author}{\bibfnamefont{A.}~\bibnamefont{Z{\"u}ttel}},
  \bibinfo{journal}{Nature} \textbf{\bibinfo{volume}{414}},
  \bibinfo{pages}{353} (\bibinfo{year}{2001}).

\bibitem[{\citenamefont{Eliashberg}(1960)}]{Eliashberg1960A}
\bibinfo{author}{\bibfnamefont{G.~M.} \bibnamefont{Eliashberg}},
  \bibinfo{journal}{Soviet Physics JETP} \textbf{\bibinfo{volume}{11}},
  \bibinfo{pages}{696} (\bibinfo{year}{1960}).

\bibitem[{\citenamefont{Szcz{\c{e}}{\'s}niak}(2006)}]{Szczesniak2006A}
\bibinfo{author}{\bibfnamefont{R.}~\bibnamefont{Szcz{\c{e}}{\'s}niak}},
  \bibinfo{journal}{Acta Physica Polonica A} \textbf{\bibinfo{volume}{109}},
  \bibinfo{pages}{179} (\bibinfo{year}{2006}).

\bibitem[{\citenamefont{Beach et~al.}(2000)\citenamefont{Beach, Gooding, and
  Marsiglio}}]{Beach2000A}
\bibinfo{author}{\bibfnamefont{K.~S.~D.} \bibnamefont{Beach}},
  \bibinfo{author}{\bibfnamefont{R.~J.} \bibnamefont{Gooding}},
  \bibnamefont{and}
  \bibinfo{author}{\bibfnamefont{F.}~\bibnamefont{Marsiglio}},
  \bibinfo{journal}{Physical Review B} \textbf{\bibinfo{volume}{61}},
  \bibinfo{pages}{5147} (\bibinfo{year}{2000}).

\bibitem[{\citenamefont{Bardeen
  et~al.}(1957{\natexlab{a}})\citenamefont{Bardeen, Cooper, and
  Schrieffer}}]{Bardeen1957A}
\bibinfo{author}{\bibfnamefont{J.}~\bibnamefont{Bardeen}},
  \bibinfo{author}{\bibfnamefont{L.~N.} \bibnamefont{Cooper}},
  \bibnamefont{and} \bibinfo{author}{\bibfnamefont{J.~R.}
  \bibnamefont{Schrieffer}}, \bibinfo{journal}{Physical Review}
  \textbf{\bibinfo{volume}{106}}, \bibinfo{pages}{162}
  (\bibinfo{year}{1957}{\natexlab{a}}).

\bibitem[{\citenamefont{Bardeen
  et~al.}(1957{\natexlab{b}})\citenamefont{Bardeen, Cooper, and
  Schrieffer}}]{Bardeen1957B}
\bibinfo{author}{\bibfnamefont{J.}~\bibnamefont{Bardeen}},
  \bibinfo{author}{\bibfnamefont{L.~N.} \bibnamefont{Cooper}},
  \bibnamefont{and} \bibinfo{author}{\bibfnamefont{J.~R.}
  \bibnamefont{Schrieffer}}, \bibinfo{journal}{Physical Review}
  \textbf{\bibinfo{volume}{108}}, \bibinfo{pages}{1175}
  (\bibinfo{year}{1957}{\natexlab{b}}).

\bibitem[{\citenamefont{Carbotte}(1990)}]{Carbotte1990A}
\bibinfo{author}{\bibfnamefont{J.~P.} \bibnamefont{Carbotte}},
  \bibinfo{journal}{Reviews of Modern Physics} \textbf{\bibinfo{volume}{62}},
  \bibinfo{pages}{1027} (\bibinfo{year}{1990}).

\bibitem[{\citenamefont{Aguilera-Navarro and
  de~Llano}(1992)}]{Aguilera-Navarro1992A}
\bibinfo{author}{\bibfnamefont{V.}~\bibnamefont{Aguilera-Navarro}}
  \bibnamefont{and} \bibinfo{author}{\bibfnamefont{M.}~\bibnamefont{de~Llano}},
  \bibinfo{journal}{Symmetries in Physics. Springer} p.~\bibinfo{pages}{28}
  (\bibinfo{year}{1992}).

\bibitem[{\citenamefont{Liu et~al.}(2019)\citenamefont{Liu, Wang, Yi, Kim, Kim,
  and Cho}}]{Liu2019A}
\bibinfo{author}{\bibfnamefont{L.}~\bibnamefont{Liu}},
  \bibinfo{author}{\bibfnamefont{C.}~\bibnamefont{Wang}},
  \bibinfo{author}{\bibfnamefont{S.}~\bibnamefont{Yi}},
  \bibinfo{author}{\bibfnamefont{K.~W.} \bibnamefont{Kim}},
  \bibinfo{author}{\bibfnamefont{J.}~\bibnamefont{Kim}}, \bibnamefont{and}
  \bibinfo{author}{\bibfnamefont{J.}~\bibnamefont{Cho}},
  \bibinfo{journal}{arXiv:1811.08548v3}  (\bibinfo{year}{2019}).

\bibitem[{\citenamefont{Chen et~al.}(2019)\citenamefont{Chen, Semenok, Troyan,
  Ivanova, Huang, Oganov, and Cui}}]{Chen2019A}
\bibinfo{author}{\bibfnamefont{W.}~\bibnamefont{Chen}},
  \bibinfo{author}{\bibfnamefont{D.~V.} \bibnamefont{Semenok}},
  \bibinfo{author}{\bibfnamefont{I.~A.} \bibnamefont{Troyan}},
  \bibinfo{author}{\bibfnamefont{A.~G.} \bibnamefont{Ivanova}},
  \bibinfo{author}{\bibfnamefont{X.}~\bibnamefont{Huang}},
  \bibinfo{author}{\bibfnamefont{A.~R.} \bibnamefont{Oganov}},
  \bibnamefont{and} \bibinfo{author}{\bibfnamefont{T.}~\bibnamefont{Cui}},
  \bibinfo{journal}{arXiv:1903.02194}  (\bibinfo{year}{2019}).

\end{thebibliography}

\end{document}